\documentclass[12pt,a4paper]{article}
\usepackage[twocolumn=false,ignoreall]{geometry}
\usepackage[final,hiresbb]{graphicx}
\usepackage[format=default,justification=raggedright,singlelinecheck=false,labelformat=simple,labelsep=quad,labelfont={bf},textfont={small},font={small}]{caption}
\usepackage{tabularx}
\usepackage{boxedminipage}
\usepackage{array}
\usepackage{dcolumn}
\usepackage{longtable}
\usepackage[alwaysadjust]{paralist}
\usepackage{parallel}
\usepackage[integers,english]{layout}
\usepackage{hyperref}
\usepackage{color}

\input{tcilatex}
\begin{document}

\begin{titlepage}
\date{March 2018}
\title{Deflection of light and time delay in closed Einstein-Straus solution}

\author{Mourad Guenouche$^{a,b}$ 
\thanks{Email: \href{mailto: guenouche$_{-}$mourad@umc.edu.dz}{guenouche$_{-}$mourad@umc.edu.dz}} \and
Sami Ryad Zouzou$^a$
\thanks{Email: \href{mailto: szouzou2000@yahoo.fr}{szouzou2000@yahoo.fr}} \\ \\
$^a$ {\small Laboratoire de Physique Th\'eorique, D\'epartement de Physique,} \\
{\small Facult\'e des Sciences Exactes, Universit\'e de Constantine 1, Algeria} \\
$^b$ {\small Universit\'e Abb\`{e}s Laghrour de Khenchela, Algeria}}

\maketitle

\begin{abstract}
We investigate strong lensing by a spherically symmetric mass distribution in the framework of the Einstein-Straus solution 
with positive cosmological constant 
and concentrate on the case of a spatially closed Universe ($k=+1$).  We develop a method based on 
integration of differential equations in order to make possible 
the computation of light deflection and time delay. By applying our results to the lensed quasar SDSS J1004+4112, we find that 
the bending of  light and time delay depend on wether the present value of the scale factor $a_0$ is or is not 
much smaller than a value close to $9.1\cdot10^{26}\unit{m}$. Beyond this value, the results are almost the 
same as for the spatially flat Universe. Moreover, it turns out that a positive cosmological constant attenuates 
the light bending in agreament with Rindler and Ishak's finding.
 \\ \\ \\ \\ \\ \\ 
 Key-Words: Cosmological parameters $\cdot$ Lensing
\end{abstract}
\end{titlepage}\setcounter{page}{2}

\section{Introduction}

On the strength of Rindler and Ishak's finding \cite{RindlerIsak}, and
several subsequent works \cite{sereno,sereno2,tschu,tschu2,mira}, it is
already well established, that a positive cosmological constant reduces the
bending of light near an isolated spherical mass in Kottler space-time.

Many authors have considered the simplest case of a spatially flat universe
(null curvature $k=0$) to investigate the same problem in a more realistic
model, namely the Einstein-Straus solution \cite{ES,Schucking} in presence
of a cosmological constant \cite{balbinot}. In a such model, it is assumed
that the bending of light by a lens happens only inside a Kottler vacuole
(Sch\"{u}cking sphere) embedded in an expanding Friedmann universe. Within
the same framework, Ishak, Rindler, Sch\"{u}cker, Kantowski et al. \cite%
{ish-rind,TSch1,kantow,TSCH}, have proved that the effect of the
cosmological constant on light bending is only diminished without however
being dropped, contrary to what has been argued in \cite{khrip,park,sim}. In
references \cite{Ishak,TD-ES,chen}, the authors have gone a step further and
investigated the cosmological constant's effect on time delay in which case
the computation of the photon's travel time outside the vacuole is
particularly simple using some properties of Euclidean geometry.

Even though the current observations predict that the universe is very close
to flat, it would nevertheless be interesting to treat the bending of light
and time delay in the Einstein-Straus solution considering the case of a
spatially closed universe (positive curvature $k=+1$) and discuss the
possible impact of positive curvature on light bending and time delay. In
such a case, the photon outside the vacuole does no longer travel in a
straight line. For this reason, the argumentation will be based only on
integration of differential equations.

We will use the same units as in reference \cite{TSch1}: astroseconds ($%
\unit{as}$), astrometers ($\unit{am}$) and astrograms ($\unit{ag}$), 
\begin{equation}
\begin{array}{l}
1\unit{as}=4.34\cdot 10^{17}\unit{s}=13.8\unit{Gyr}, \\ 
1\unit{am}=1.30\cdot 10^{26}\unit{m}=4221\unit{Mpc}, \\ 
1\unit{ag}=6.99\cdot 10^{51}\unit{kg}=3.52\cdot 10^{21}M_{\odot },%
\end{array}
\label{units}
\end{equation}%
where $M_{\odot }$ is the solar mass. In these units 
\begin{equation}
c=1\unit{am}\unit{as}^{-1},\quad 8\pi G=1\unit{am}^{3}\unit{as}^{-2}\unit{ag}%
^{-1},\quad H_{0}=1\unit{as}^{-1},
\end{equation}%
where $H_{0}$ is the Hubble constant.

\section{Matching of Kottler's and closed Friedmann's solutions with a
cosmological constant}

In order to construct the closed Einstein-Straus metric, we shall start by
matching between the Kottler and closed Friedmann metrics on the Sch\"{u}%
cking sphere (Kottler vacuole) by calculating the Jacobian of the
transformation between the Schwarzschild and the Friedmann coordinates. Let
us denote by $(T,r,\theta ,\varphi )$ the Schwarzschild coordinates and by $%
(t,\chi ,\theta ,\varphi )$ the Friedmann coordinates. The Kottler metric 
\begin{equation}
ds^{2}=B(r)dT^{2}-B(r)^{-1}dr^{2}-r^{2}(d\theta ^{2}+\sin ^{2}\theta
d\varphi ^{2}),\quad B(r):=1-\dfrac{M}{4\pi r}-\dfrac{\Lambda }{3}r^{2},
\label{kottmetric}
\end{equation}%
reigns inside a vacuole of radius $r_{Sch\ddot{u}}(T)$ centered around a
spherical mass distribution (the lens) with $r\leq r_{Sch\ddot{u}}$. The
Friedmann metric in the case of a spatial positive curvature, $k=+1$, 
\begin{equation}
ds^{2}=dt^{2}-a(t)^{2}[d\chi ^{2}+\sin ^{2}\chi (d\theta ^{2}+\sin
^{2}\theta d\varphi ^{2})],  \label{closed-fried-metric}
\end{equation}%
describes the space-time geometry outside the vacuole, $\chi \geq \chi _{Sch%
\ddot{u}}$, where the scale factor $a(t)$ is determined by integration of
the Friedmann equation 
\begin{equation}
\dfrac{da}{dt}=f(a),\quad f(a):=\sqrt{\frac{A}{a}+\frac{\Lambda }{3}a^{2}-k},
\label{friedeqt}
\end{equation}%
where the constant $A$ results from the energy conservation law 
\begin{equation}
A:=\rho \,a^{3}/3=\rho _{dust0}a_{0}^{3}/3,
\end{equation}%
for a non-relativistic matter-dominated universe with the present dust
density $\rho _{dust0}$, 
\begin{equation}
\rho _{dust0}=3-\Lambda +3\Omega _{k0},\quad \Omega _{k0}:=ka_{0}^{-2},
\end{equation}%
computed using the fact that the Hubble constant $H_{0}=a_{t}(0)/a_{0}$ in
the system of units (\ref{units}) is $1\unit{as}{}^{-1}$ (the lower indice $%
t $ denotes differentiation with respect to time). The parameters $\Omega
_{k0} $ and $a_{0}$ represent respectively the curvature density of space at
the present time and the scale factor at the present time. The two solutions
are connected at the boundary of the Sch\"{u}cking sphere by requiring 
\begin{equation}
r_{Sch\ddot{u}}(T):=a(t)\sin \chi _{Sch\ddot{u}}.  \label{matchcond}
\end{equation}%
The mass $M$ is expressed in terms of the Sch\"{u}cking radius as 
\begin{equation}
M=\dfrac{4\pi }{3}r_{Sch\ddot{u}}^{3}\rho =4\pi A\sin ^{3}\chi _{Sch\ddot{u}%
}\,,
\end{equation}%
which can be inverted to give 
\begin{equation}
\chi _{Sch\ddot{u}}=\arcsin \left[ \left( \frac{M}{4\pi A}\right) ^{1/3}%
\right] .  \label{Sch-radius}
\end{equation}%
So we have on the Sch\"{u}cking sphere 
\begin{equation}
B_{Sch\ddot{u}}:=B(r_{Sch\ddot{u}})=1-\left( \dfrac{A}{a}+\dfrac{\Lambda }{3}%
a^{2}\right) \sin \chi _{Sch\ddot{u}}^{2}.  \label{Bsch}
\end{equation}%
To connect the two solutions, proceeding in an analogous way to what has
been done by Balbinot and Sch\"{u}cker \cite{balbinot,TSch1}, we will pass
from Schwarzschild coordinates $(T,r)$ and Friedmann coordinates $(t,\chi )$
to the new coordinate system $(b,r)$. In this new coordinante system, the
Kottler metric can be rewritten as 
\begin{equation}
ds^{2}=B\Psi (b)^{2}db^{2}-\frac{1}{B}dr^{2}-r^{2}d\Omega ^{2}.
\end{equation}%
where the function $\Psi (b)$ is defined by 
\begin{equation}
\Psi (b):=\frac{dT}{db}.
\end{equation}%
We now rewrite first the Friedmann metric in the coordinate system $(a,\chi
) $, 
\begin{equation}
ds^{2}=\frac{da^{2}}{f(a)^{2}}-a^{2}d\chi ^{2}-a^{2}\sin ^{2}\chi d\Omega
^{2}.
\end{equation}%
and convert the factor $a^{2}\sin ^{2}\chi $ in $r^{2}$ under a second
coordinate transformation $(a,\chi )\rightarrow (b,r)$, 
\begin{equation}
a:=\Phi (b,r),\quad \sin \chi :=r/\Phi (b,r),
\end{equation}%
with the boundary condition that at the Sch\"{u}cking radius, old and new
time coordinates coincide, 
\begin{equation}
a=b=\Phi (b,b\sin \chi _{Sch\ddot{u}}).  \label{Boundary condition}
\end{equation}%
The Friedmann metric then becomes 
\begin{eqnarray}
ds^{2} &=&\Phi _{b}^{2}\left( \frac{1}{C_{1}^{2}}-\frac{r^{2}}{\Phi
^{2}-r^{2}}\right) db^{2}-\left[ \frac{(\Phi -r\Phi _{r})\,^{2}}{\Phi
^{2}-r^{2}}-\frac{\Phi _{r}^{2}}{C_{1}^{2}}\right] dr^{2}  \nonumber \\
&&+2\,\Phi _{b}\left[ \frac{r(\Phi -r\Phi _{r})}{\Phi ^{2}-r^{2}}+\frac{\Phi
_{r}}{C_{1}^{2}}\,\right] dbdr-r^{2}d\Omega ^{2},
\end{eqnarray}%
with $\Phi _{b}=\partial \Phi /\partial b$, $\Phi _{r}=\partial \Phi
/\partial r$ and 
\begin{equation}
C_{1}:=\sqrt{\frac{A}{\Phi }+\frac{\Lambda }{3}\Phi ^{2}-1}.
\end{equation}%
Since the metric should be diagonal, this implies the absence of mixed
terms, i.e. $g_{br}^{F}=0$. Then 
\begin{equation}
\Phi _{r}=-\,\dfrac{rC_{1}^{2}}{\Phi B_{1}}\,,\quad B_{1}:=1-\left( \dfrac{A%
}{\Phi ^{3}}+\dfrac{\Lambda }{3}\right) r^{2}.
\end{equation}%
Therefore the Friedmann metric can be put in the form\ 
\begin{equation}
ds^{2}=\frac{\Phi _{b}^{2}}{1-r^{2}/\Phi ^{2}}\frac{B_{1}}{C_{1}^{2}}%
\!\,\,db^{2}-\frac{1}{B_{1}}dr^{2}-r^{2}d\Omega ^{2}.  \label{friednew}
\end{equation}%
Differentiating the boundary condition (\ref{Boundary condition}) with
respect to $b$, we obtain at $\chi =\chi _{Sch\ddot{u}}$ 
\begin{equation}
\Phi _{b}|_{Sch\ddot{u}}:=\Phi _{b}(b,b\sin \chi _{Sch\ddot{u}})=1-\Phi
_{r}|_{Sch\ddot{u}}\sin \chi _{Sch\ddot{u}}=\frac{\cos ^{2}\chi _{Sch\ddot{u}%
}}{B_{Sch\ddot{u}}}.
\end{equation}%
The matching of the two solutions continuously on the Sch\"{u}cking sphere
in this coordinate system $(b,r,\theta ,\varphi )$ results in 
\begin{equation}
\left. g_{bb}^{F}\right\vert _{Sch\ddot{u}}=\left. g_{bb}^{K}\right\vert
_{Sch\ddot{u}},\quad \left. g_{rr}^{F}\right\vert _{Sch\ddot{u}}=\left.
g_{rr}^{K}\right\vert _{Sch\ddot{u}}.  \label{matching}
\end{equation}%
We can demonstrate that 
\begin{equation}
\left. B_{1}\right\vert _{Sch\ddot{u}}=B_{Sch\ddot{u}},\quad \left.
C_{1}\right\vert _{Sch\ddot{u}}=\dfrac{C_{Sch\ddot{u}}}{\tan \chi _{Sch\ddot{%
u}}},
\end{equation}%
with $C_{Sch\ddot{u}}$ defined by 
\begin{equation}
C_{Sch\ddot{u}}:=\sqrt{1-\frac{B_{Sch\ddot{u}}}{\cos ^{2}\chi _{Sch\ddot{u}}}%
}.
\end{equation}%
The relations (\ref{matching}) yield 
\begin{equation}
\Psi (b)=\dfrac{\sin \chi _{Sch\ddot{u}}}{B_{Sch\ddot{u}}(b)C_{Sch\ddot{u}%
}(b)}.
\end{equation}%
Repeated use of the chain rule then gives 
\begin{equation}
\begin{array}{ll}
\dfrac{\partial t}{\partial T}=\dfrac{\partial t}{\partial a}\dfrac{\partial
\Phi }{\partial b}\dfrac{\partial b}{\partial T}, & \dfrac{\partial t}{%
\partial r}=\dfrac{\partial t}{\partial a}\dfrac{\partial \Phi }{\partial r},
\\ 
\dfrac{\partial \chi }{\partial T}=\dfrac{\partial \chi }{\partial b}\dfrac{%
\partial b}{\partial T}, & \dfrac{\partial \chi }{\partial r}=\dfrac{1}{\cos
\chi }\left( \dfrac{1}{\Phi }-\dfrac{r}{\Phi ^{2}}\dfrac{\partial \Phi }{%
\partial r}\right) .%
\end{array}%
\end{equation}%
Hence, the Jacobian of the coordinate transformation $(T,r)\rightarrow
(t,\chi )$ at the Sch\"{u}cking radius $\chi =\chi _{Sch\ddot{u}}$ is given
by 
\begin{equation}
\begin{array}{ll}
\left. \dfrac{\partial t}{\partial T}\right\vert _{Sch\ddot{u}}=\cos \chi
_{Sch\ddot{u}}, & \left. \dfrac{\partial t}{\partial r}\right\vert _{Sch%
\ddot{u}}=-\dfrac{C_{Sch\ddot{u}}\cos \chi _{Sch\ddot{u}}}{B_{Sch\ddot{u}}},
\\ 
\left. \dfrac{\partial \chi }{\partial T}\right\vert _{Sch\ddot{u}}=-\dfrac{%
C_{Sch\ddot{u}}\cos \chi _{Sch\ddot{u}}}{a}, & \left. \dfrac{\partial \chi }{%
\partial r}\right\vert _{Sch\ddot{u}}=\dfrac{\cos \chi _{Sch\ddot{u}}}{%
aB_{Sch\ddot{u}}},%
\end{array}
\label{Jacob}
\end{equation}%
and the Jacobian of the inverse coordinate transformation $\left( t,\chi
\right) \rightarrow \left( T,r\right) $ is given by 
\begin{equation}
\begin{array}{ll}
\left. \dfrac{\partial T}{\partial t}\right\vert _{Sch\ddot{u}}=\dfrac{\cos
\chi _{Sch\ddot{u}}}{B_{Sch\ddot{u}}}, & \left. \dfrac{\partial T}{\partial
\chi }\right\vert _{Sch\ddot{u}}=\dfrac{aC_{Sch\ddot{u}}\cos \chi _{Sch\ddot{%
u}}}{B_{Sch\ddot{u}}}, \\ 
\left. \dfrac{\partial r}{\partial t}\right\vert _{Sch\ddot{u}}=C_{Sch\ddot{u%
}}\cos \chi _{Sch\ddot{u}}, & \left. \dfrac{\partial r}{\partial \chi }%
\right\vert _{Sch\ddot{u}}=a\cos \chi _{Sch\ddot{u}}.%
\end{array}
\label{inverseJacob}
\end{equation}

To compare the Schwarzschild coordinate time $T$ with that of Friedmann $t$
on the Sch\"{u}cking sphere $\chi =\chi _{Sch\ddot{u}}$, we consider the
parameterized curve, $T=p$, $r=b\sin \chi _{Sch\ddot{u}}$ ($\theta =\pi /2$, 
$\varphi =0$). Its 4-velocity is given by 
\begin{equation}
\begin{array}{l}
\dfrac{dT}{dp}=1, \\ 
\dfrac{dr}{dp}=\dfrac{dr}{db}\left. \dfrac{db}{dT}\right\vert _{Sch\ddot{u}}%
\dfrac{dT}{dp}=B_{Sch\ddot{u}}C_{Sch\ddot{u}},%
\end{array}%
\end{equation}%
in Schwarzschild coordinates and, 
\begin{equation}
\begin{array}{l}
\dfrac{dt}{dp}=\left. \dfrac{\partial t}{\partial T}\right\vert _{Sch\ddot{u}%
}\dfrac{dT}{dp}+\left. \dfrac{\partial t}{\partial r}\right\vert _{Sch\ddot{u%
}}\dfrac{dr}{dp}=\dfrac{B_{Sch\ddot{u}}}{\cos \chi _{Sch\ddot{u}}}, \\ 
\dfrac{d\chi }{dp}=\left. \dfrac{\partial \chi }{\partial T}\right\vert _{Sch%
\ddot{u}}\dfrac{dT}{dp}+\left. \dfrac{\partial \chi }{\partial r}%
\!\right\vert _{Sch\ddot{u}}\dfrac{dr}{dp}=0,%
\end{array}%
\end{equation}%
in Friedmann coordinates. Finally, we deduce the relation 
\begin{equation}
\left. \dfrac{dT}{dt}\right\vert _{Sch\ddot{u}}=\dfrac{dT}{dp}\,\dfrac{dp}{dt%
}=\frac{\cos \chi _{Sch\ddot{u}}}{B_{Sch\ddot{u}}},  \label{match-cond}
\end{equation}%
which allows us to pass from the Schwarzschild coordinate time to the
Friedmann coordinate time and vice versa.

\section{Equations of null geodesic motion}

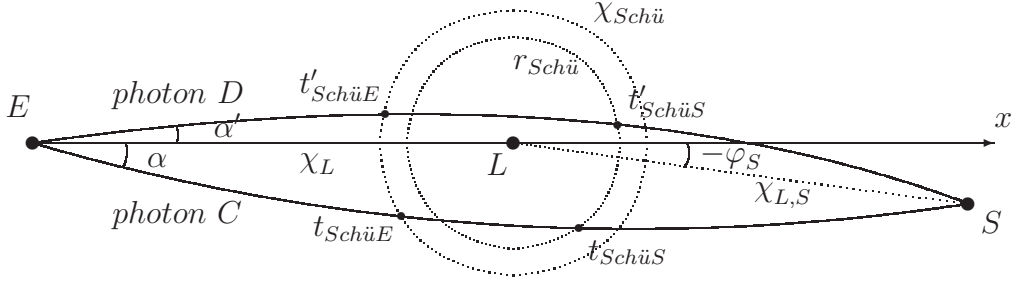
\begin{figure}[htbp] \centering%
\begin{picture}(400,150)
\put(20,70){\vector(1,0){360}}
\put(380,76){$x$}
\put(10,80){$E$}
\put(20,70){\circle*{5}}
\qbezier[80](200,70)(265,61)(370,47)
\put(375,36){$S$}
\qbezier(264,61)(267,65)(265,70)
\put(270,63){$-\varphi_S$}
\put(370,47){\circle*{5}}
\put(200,70){\circle*{5}}
\put(190,58){$L$}
\qbezier[10000](20,70)(230,100)(370,47)
\qbezier[10000](20,70)(190,20)(370,47)
\put(88,72.4){$\alpha^{\prime}$}
\qbezier(74,70)(75,73)(74,76)
\put(63,61){$\alpha$}
\qbezier(55,70)(56,65)(54,61)
\put(50,85){$photon$ $D$}
\put(50,40){$photon$ $C$}
\put(120,61){$\chi_L$}
\put(290,50){$\chi_{L,S}$}

\qbezier[40](200,120)(246,118)(250,70)
\qbezier[40](200,120)(154,118)(150,70)
\qbezier[40](200,20)(153,24)(150,70)
\qbezier[40](200,20)(248,24)(250,70)

\qbezier[40](200,110)(236,108)(240,70)
\qbezier[40](200,110)(164,108)(160,70)
\qbezier[40](200,30)(163,34)(160,70)
\qbezier[40](200,30)(238,34)(240,70)

\put(239,77){\circle*{3}}
\put(152,81){\circle*{3}}
\put(224.5,38){\circle*{3}}
\put(158,42.5){\circle*{3}}

\put(230,118){$\chi_{Sch\ddot{u}}$}
\put(200,98){$r_{Sch\ddot{u}}$}

\put(243,83){$t_{Sch\ddot{u}S}^{\prime}$}

\put(228,26){$t_{Sch\ddot{u}S}$}

\put(120,89){$t_{Sch\ddot{u}E}^{\prime}$}

\put(125,35){$t_{Sch\ddot{u}E}$}

\end{picture}%
\caption{Two photons emitted by a source $S$, bent inside the
Schücking sphere and finally received at Earth $E$. Outside the Schücking
sphere these two photons do not travel in straight lines}\label{def}%
\end{figure}

Since these are the final conditions on Earth that are known, it seems more
convenient to consider two photons $C$ and $D$ emitted by a source $S$
(quasar) at different times $t_{S}\neq t_{S}^{\prime }$, follow different
trajectories and received on Earth $E$ simultaneously at $%
t_{E}=t_{E}^{\prime }=0$ with the angles $\alpha $ and $\alpha ^{\prime }$.
As shown in Fig. \ref{def}, these two photons, during their propagation in
curved Friedmann's space-time, penetrate the Sch\"{u}cking sphere,
respectively at $t_{Sch\ddot{u}S}$ and $t_{Sch\ddot{u}S}^{\prime }$, where
they get deflected by a lens $L$ (galaxy cluster) respectively at minimum
distances $r_{P}$ and $r_{P}^{\prime }$ (peri-lens), which are much larger
than the Schwarzschild radius ($r_{Schw}:=2GM=M/4\pi $), and then leave the
Sch\"{u}cking sphere respectively at $t_{Sch\ddot{u}E}$ and $t_{Sch\ddot{u}%
E}^{\prime }$.

In the presence of a cosmological constant and a nonzero spatial curvature,
exact analytical solutions of the Friedmann equation (\ref{friedeqt}) for
the cosmic time $t(a)$ can be obtained in terms of elliptic integrals of
second and third kind whose inverse function for the scale factor $a(t)$ is
not known. Nevertheless, one can alternatively proceed to a numerical
resolution. We will therefore determine the scale factor $a(t)$ by numerical
integration of the Friedmann equation with final condition $a(t=0)=a_{0}$.
In order to compare our results with those obtained previously in the flat
case, we use the same experimentally measured value of cosmological constant 
$\Lambda =0.77\cdot 3$ $\unit{am}{}^{-2}\pm 20\%$ or in $\unit{cm}{}^{-2}$
units $\Lambda =1.36$ $\cdot 10^{-56}\unit{cm}{}^{-2}\pm 20\%$ as in
references \cite{tschu2,TSch1,TD-ES} and dust density $\rho
_{dust0}=3-\Lambda +3a_{0}^{-2}$.

In order to determine the Earth-Lens geodesic distance $\chi _{L}$ and the
Earth-Source geodesic distance $\chi _{S}$, we will also need to solve
numerically the radial null geodesic 
\begin{equation}
d\chi =-\frac{dt}{a(t)},
\end{equation}%
which ensures that $\chi (t)$ is a decreasing function of time. But, the
right-hand side of this equation depends on the solution $a(t)$ of Friedmann
equation which is not known. We can get round this problem by introducing
the function $f(a)$ through the Friedmann equation $dt=da/f(a)$ and the
redshift formula $1+z=a_{0}/a$, 
\begin{equation}
\chi (z)=\int_{\tfrac{a_{0}}{1+z}}^{a_{0}}\frac{da}{af(a)},  \label{distES}
\end{equation}%
where we have taken the origin $\chi (z=0)=0$ at the position of the Earth,
thereby $\chi _{L}$ and $\chi _{S}$ will be respectively calculated for any
given value of $z_{L}$ and $z_{S}$. Since these are the final conditions at
the arrival on Earth which are given, it is necessary to proceed backwards
in time. We will first determine $t_{Sch\ddot{u}E}^{\prime }$ and $t_{Sch%
\ddot{u}E}$, then $t_{Sch\ddot{u}S}^{\prime }$ and $t_{Sch\ddot{u}S}$ to
calculate the angle $\varphi _{S}$ and finally $t_{S}^{\prime }$ and $t_{S}$
for the computation of time delay.

\subsection{Null geodesics between the Sch\"{u}cking sphere and the Earth}

\label{1}In this region, the photon trajectory is governed by the closed
Friedmann metric (\ref{closed-fried-metric}). The corresponding nonzero
Christoffel symbols in the equatorial plane $\theta =\pi /2$ are 
\begin{equation}
\begin{array}{lll}
\Gamma _{\chi \chi }^{t}=aa_{t}, & \Gamma _{\varphi \varphi }^{t}=aa_{t}\sin
^{2}\chi , & \Gamma _{\varphi \varphi }^{\chi }=-\sin \chi \cos \chi , \\ 
\Gamma _{t\chi }^{\chi }=a_{t}/a, & \Gamma _{t\varphi }^{\varphi }=a_{t}/a,
& \Gamma _{\chi \varphi }^{\varphi }=\cot \chi .%
\end{array}%
\end{equation}%
Then the geodesic equations read 
\begin{eqnarray}
&&\ddot{t}+aa_{t}(\dot{\chi}^{2}+\sin ^{2}\chi \dot{\varphi}^{2})=0,
\label{feqt1} \\
&&\ddot{\chi}+2a^{-1}a_{t}\dot{t}\dot{\chi}-\sin \chi \cos \chi \dot{\varphi}%
^{2}=0, \\
&&\ddot{\varphi}+2(a^{-1}a_{t}\dot{t}+\cot \chi \dot{\chi})\dot{\varphi}=0,
\label{feqt3}
\end{eqnarray}%
where $\dot{~}:=d/dp$, with $p$ an affine parameter other than $s$ because
the spacetime interval for light is zero $ds=0$. It should be noted that the
second equation might not be needed for our purpose.

The upper photon would arrive at Earth with final conditions ($p=0$)%
\begin{equation}
\begin{array}{lll}
t=0, & \chi =\chi _{L}, & \varphi =\pi , \\ 
\dot{t}=1, & \dot{\chi}=\dfrac{1}{a_{0}\sqrt{1+\cos ^{2}\chi _{L}\tan
^{2}\alpha ^{\prime }}}, & \dot{\varphi}=\dfrac{\cot \chi _{L}\tan \alpha
^{\prime }}{a_{0}\sqrt{1+\cos ^{2}\chi _{L}\tan ^{2}\alpha ^{\prime }}},%
\end{array}%
\end{equation}%
where we use the fact that the physical angle $\alpha ^{\prime }$ coincides
with the coordinate angle $\arctan |\tan \chi _{L}\dot{\varphi}/\dot{\chi}|$%
. These final conditions make it possible to integrate the geodesic
equations which yield 
\begin{equation}
\dot{t}=\frac{a_{0}}{a(t)},\quad \dot{\varphi}=\dfrac{a_{0}\chi _{P}^{\prime
}}{a^{2}\sin ^{2}\chi },\quad \varphi =\pi -\arcsin \dfrac{\chi _{P}^{\prime
}\cot \chi }{\sqrt{1-\chi _{P}^{\prime 2}}}+\beta ^{\prime },  \label{sol}
\end{equation}%
where the constants $\chi _{P}^{\prime }$ (peri-lens) and $\beta ^{\prime }$
are defined by 
\begin{equation}
\chi _{P}^{\prime }:=\dfrac{\cos \chi _{L}\sin \chi _{L}\tan \alpha ^{\prime
}}{\sqrt{1+\cos ^{2}\chi _{L}\tan ^{2}\alpha ^{\prime }}},\quad \beta
^{\prime }:=\arcsin \dfrac{\chi _{P}^{\prime }\cot \chi _{L}}{\sqrt{1-\chi
_{P}^{\prime 2}}}.  \label{perilens1}
\end{equation}

In references \cite{TSch1,TD-ES}, the distance traveled by the photon
between its exit point from the vacuole and its arrival point on Earth,
noted by $\chi _{E,Sch\ddot{u}E}^{\prime }$ (or $\chi _{E,Sch\ddot{u}E}$),
is computed by means of some properties of Euclidean geometry which are no
longer valid in a curved Friedmann space-time where the geodesic distances
are not straight lines. The same remark must be made regarding the distance
traveled by the photon between the source and its entry point into the
vacuole ($\chi _{S,Sch\ddot{u}S}^{\prime }$ and $\chi _{S,Sch\ddot{u}S}$).
For this reason, we have developed a method based only on Friedmann
differential equations which remains valid either in flat or curved
space-time.

Differentiating the solution $\varphi $ (\ref{sol}) with respect to $\chi $
and substituting into the closed Friedmann metric (\ref{closed-fried-metric}%
) for a photon, we can easily get an equation linking the variable $\chi $
with time 
\begin{equation}
\int \frac{dt}{a(t)}=\int \frac{d\chi }{g^{\prime }(\chi )},\quad g^{\prime
}(\chi ):=\sqrt{1-\chi _{P}^{\prime 2}/\sin ^{2}\chi },  \label{varpchi}
\end{equation}%
where we have taken into account that $\chi $ between the vacuole and the
Earth increases with time. The right-hand side of this equation could be
easily evaluated between $\chi _{Sch\ddot{u}}$ and $\chi _{L}$ to give
precisely the geodesic distance $\chi _{E,Sch\ddot{u}E}^{\prime }$ in terms
of the $\arcsin $ function, but we do not know the expression of scale
factor in terms of time in the left-hand side, so we shall first have to
integrate with respect to the scale factor by inserting the function $f(a)$
via $dt=da/f(a)$. We obtain 
\begin{equation}
\int_{a_{Sch\ddot{u}E}^{\prime }}^{a_{0}}\frac{da}{af(a)}=\arcsin \frac{\cos
\chi _{Sch\ddot{u}}}{\sqrt{1-\chi _{P}^{\prime 2}}}-\arcsin \frac{\cos \chi
_{L}}{\sqrt{1-\chi _{P}^{\prime 2}}},  \label{form1}
\end{equation}%
from which we easily deduce numerically the value of $a_{Sch\ddot{u}%
E}^{\prime }$ where $\chi _{Sch\ddot{u}}$ and $\chi _{L}$ are respectively
given by (\ref{Sch-radius}) and (\ref{distES}). The value of $t_{Sch\ddot{u}%
E}^{\prime }$, at which the upper photon emerges from the Sch\"{u}cking
sphere, is then calculated by numerical integration of Friedmann equation,
i.e. 
\begin{equation}
t_{Sch\ddot{u}E}^{\prime }=\int_{a_{0}}^{a_{Sch\ddot{u}E}^{\prime }}\dfrac{da%
}{f(a)}.  \label{form2}
\end{equation}%
We point out that an equation linking the variable $\varphi $ with time may
be established by the same manner as before and serve to deduce the value of 
$t_{Sch\ddot{u}E}^{\prime }$ using the polar angle at the exit point from
the vacuole 
\begin{equation}
\varphi _{Sch\ddot{u}E}^{\prime }=\pi -\arcsin \dfrac{\chi _{P}^{\prime
}\cot \chi _{Sch\ddot{u}}}{\sqrt{1-\chi _{P}^{\prime 2}}}+\beta ^{\prime }.
\label{varphiup}
\end{equation}

Similar formulae apply in the case of the lower trajectory photon, with $\pi 
$ replaced by $-\pi $ and $\alpha ^{\prime }$ replaced by $-\alpha $. The
final conditions upon arrival on Earth are 
\begin{equation}
\begin{array}{lll}
t=0, & \chi =\chi _{L}, & \varphi =-\pi , \\ 
\dot{t}=1, & \dot{\chi}=\dfrac{1}{a_{0}\sqrt{1+\cos ^{2}\chi _{L}\tan
^{2}\alpha }}, & \dot{\varphi}=\dfrac{-\cot \chi _{L}\tan \alpha }{a_{0}%
\sqrt{1+\cos ^{2}\chi _{L}\tan ^{2}\alpha }}.%
\end{array}%
\end{equation}%
The integration of geodesic equations thus yields 
\begin{equation}
\dot{t}=\frac{a_{0}}{a(t)},\quad \dot{\varphi}=\dfrac{-a_{0}\chi _{P}}{%
a^{2}\sin ^{2}\chi },\quad \varphi =-\pi +\arcsin \dfrac{\chi _{P}\cot \chi 
}{\sqrt{1-\chi _{P}^{2}}}-\beta ,
\end{equation}%
with 
\begin{equation}
\chi _{P}:=\dfrac{\cos \chi _{L}\sin \chi _{L}\tan \alpha }{\sqrt{1+\cos
^{2}\chi _{L}\tan ^{2}\alpha }},\quad \beta :=\arcsin \dfrac{\chi _{P}\cot
\chi _{L}}{\sqrt{1-\chi _{P}^{2}}}.  \label{perilens2}
\end{equation}%
It follows that the polar angle at which the lower trajectory photon emerges
from the Sch\"{u}cking sphere is 
\begin{equation}
\varphi _{Sch\ddot{u}E}=-\pi +\arcsin \dfrac{\chi _{P}\cot \chi _{Sch\ddot{u}%
}}{\sqrt{1-\chi _{P}^{2}}}-\beta .  \label{varphilow}
\end{equation}

The value of $t_{Sch\ddot{u}E}$ at which the lower photon emerges from the
Sch\"{u}cking sphere is obtained by means of a formula similar to (\ref%
{form2}), i.e. 
\begin{equation}
t_{Sch\ddot{u}E}=\int_{a_{0}}^{a_{Sch\ddot{u}E}}\dfrac{da}{f(a)},
\label{fdiffeqt}
\end{equation}%
in terms of $a_{Sch\ddot{u}E}$, which in turn is calculated by means of a
formula similar to (\ref{form1}), i.e.%
\begin{equation}
\int_{a_{Sch\ddot{u}E}}^{a_{0}}\frac{da}{af(a)}=\arcsin \frac{\cos \chi _{Sch%
\ddot{u}}}{\sqrt{1-\chi _{P}^{2}}}-\arcsin \frac{\cos \chi _{L}}{\sqrt{%
1-\chi _{P}^{2}}}.  \label{form3}
\end{equation}

However, as in the flat case \cite{TD-ES}, one can proceed otherwise,
without further numerical computation to obtain the time $t_{Sch\ddot{u}E}$
for lower trajectory photon. The method consists in determining $t_{Sch\ddot{%
u}E}$ by difference with $t_{Sch\ddot{u}E}^{\prime }$ through an approximate
analytical expression. Combining (\ref{form1}) and (\ref{form3}), one gets,
after evaluating the integrals with respect to time, 
\begin{eqnarray}
\int_{t_{Sch\ddot{u}E}}^{t_{Sch\ddot{u}E}^{\prime }}\frac{dt}{a(t)}
&=&\arcsin \frac{\cos \chi _{Sch\ddot{u}}}{\sqrt{1-\chi _{P}^{2}}}-\arcsin 
\frac{\cos \chi _{L}}{\sqrt{1-\chi _{P}^{2}}}-  \nonumber \\
&&\arcsin \frac{\cos \chi _{Sch\ddot{u}}}{\sqrt{1-\chi _{P}^{\prime 2}}}%
+\arcsin \frac{\cos \chi _{L}}{\sqrt{1-\chi _{P}^{\prime 2}}}.
\end{eqnarray}%
To approximate the left-hand side of this equation, one may use the fact
that the scale factor $a(t)$ varies significantly only on cosmological time
scales. Then 
\begin{equation}
\text{LHS}\simeq \frac{t_{Sch\ddot{u}E}^{\prime }-t_{Sch\ddot{u}E}}{a_{Sch%
\ddot{u}E}^{\prime }}.
\end{equation}%
A plausible approximation could be done for the right-hand side in the limit
of small peri-lens, $\chi _{P}\ll 1\ $and $\chi _{P}^{\prime }\ll 1$ (of
order $10^{-6}$ in our case), which is motivated by the observed small $%
\alpha $ and $\alpha ^{\prime }$ of the order of a few arcseconds ($\sim
10^{-5}$). One gets up to second order in the physical angles $\alpha $ and $%
\alpha ^{\prime }$, on account of (\ref{perilens1}) and (\ref{perilens2}), 
\begin{eqnarray}
\text{RHS} &\simeq &\frac{1}{2}(\cot \chi _{Sch\ddot{u}}-\cot \chi
_{L})(\chi _{P}^{2}-\chi _{P}^{\prime 2}) \\
&\simeq &\frac{1}{2}\sin (\chi _{L}-\chi _{Sch\ddot{u}})\cos ^{2}\chi _{L}%
\frac{\sin \chi _{L}}{\sin \chi _{Sch\ddot{u}}}(\alpha ^{2}-\alpha ^{\prime
2}).
\end{eqnarray}%
Equating the two sides, we get 
\begin{equation}
t_{Sch\ddot{u}E}^{\prime }-t_{Sch\ddot{u}E}\simeq \frac{1}{2}a_{Sch\ddot{u}%
E}^{\prime }\sin (\chi _{L}-\chi _{Sch\ddot{u}})\cos ^{2}\chi _{L}\frac{\sin
\chi _{L}}{\sin \chi _{Sch\ddot{u}}}(\alpha ^{2}-\alpha ^{\prime 2}).
\label{difexit}
\end{equation}%
Knowing the value of $t_{Sch\ddot{u}E}^{\prime }$ from (\ref{form2}), one
can deduce the value of $t_{Sch\ddot{u}E}$\ from this approximate
expression. This is clearly positive for $\alpha >\alpha ^{\prime }$,
meaning that the lower photon leaves the vacuole before the upper one.

The two 4-velocities of the upper and lower photons at points of exit from
the vacuole are respectively, 
\begin{equation}
\dot{t}_{Sch\ddot{u}E}^{\prime }=\frac{a_{0}}{a_{Sch\ddot{u}E}^{\prime }}%
,\quad \dot{\chi}_{Sch\ddot{u}E}^{\prime }=\dfrac{a_{0}g{}_{Sch\ddot{u}%
}^{\prime }}{a_{Sch\ddot{u}E}^{\prime 2}},\quad \dot{\varphi}_{Sch\ddot{u}%
E}^{\prime }=\dfrac{a_{0}\chi _{P}^{\prime }}{r_{Sch\ddot{u}E}^{\prime 2}},
\end{equation}%
and 
\begin{equation}
\dot{t}_{Sch\ddot{u}E}=\frac{a_{0}}{a_{Sch\ddot{u}E}},\quad \dot{\chi}_{Sch%
\ddot{u}E}=\dfrac{a_{0}g{}_{Sch\ddot{u}}}{a_{Sch\ddot{u}E}^{2}},\quad \dot{%
\varphi}_{Sch\ddot{u}E}=\dfrac{-a_{0}\chi _{P}}{r_{Sch\ddot{u}E}^{2}},
\end{equation}%
using (\ref{sol}), in Friedmann coordinates, where the value of $a_{Sch\ddot{%
u}E}$ is calculated by numerical integration of Friedmann equation (\ref%
{fdiffeqt}), with $g{}_{Sch\ddot{u}}^{\prime }:=g^{\prime }(\chi _{Sch\ddot{u%
}})$, $g{}_{Sch\ddot{u}}:=g(\chi _{Sch\ddot{u}})$ (\ref{varpchi}), and $%
r_{Sch\ddot{u}E}^{\prime }=a_{Sch\ddot{u}E}^{\prime }\sin \chi _{Sch\ddot{u}%
} $, $r_{Sch\ddot{u}E}=a_{Sch\ddot{u}E}\sin \chi _{Sch\ddot{u}}$, using the
matching condition (\ref{matchcond}). These two 4-velocities can be
translated, thanks to the inverse Jacobian (\ref{inverseJacob}), into
Schwarzschild coordinates, 
\begin{equation}
\dot{T}_{Sch\ddot{u}E}^{\prime }=\frac{a_{0}\cos \chi _{Sch\ddot{u}}}{a_{Sch%
\ddot{u}E}^{\prime }B_{Sch\ddot{u}E}^{\prime }}(1+C_{Sch\ddot{u}E}^{\prime
}g{}_{Sch\ddot{u}}^{\prime }),\quad \dot{r}_{Sch\ddot{u}E}^{\prime }=\frac{%
a_{0}\cos \chi _{Sch\ddot{u}}}{a_{Sch\ddot{u}E}^{\prime }}(C_{Sch\ddot{u}%
E}^{\prime }+g{}_{Sch\ddot{u}}^{\prime }),
\end{equation}%
for the upper photon, and 
\begin{equation}
\dot{T}_{Sch\ddot{u}E}=\dfrac{a_{0}\cos \chi _{Sch\ddot{u}}}{a_{Sch\ddot{u}%
E}^{\prime }B_{Sch\ddot{u}E}^{\prime }}(1+C_{Sch\ddot{u}E}g{}_{Sch\ddot{u}%
}),\quad \dot{r}_{Sch\ddot{u}E}=\dfrac{a_{0}\cos \chi _{Sch\ddot{u}}}{a_{Sch%
\ddot{u}E}}(C_{Sch\ddot{u}E}+g{}_{Sch\ddot{u}}),
\end{equation}%
for the lower photon, where $B_{Sch\ddot{u}E}^{\prime }$ and $B_{Sch\ddot{u}%
E}$ have been defined in (\ref{Bsch}) with $C_{Sch\ddot{u}E}^{\prime }=\sqrt{%
1-B_{Sch\ddot{u}E}^{\prime }/\cos ^{2}\chi _{Sch\ddot{u}}}$ and $C_{Sch\ddot{%
u}E}=\sqrt{1-B_{Sch\ddot{u}E}/\cos ^{2}\chi _{Sch\ddot{u}}}$.

Let $\gamma _{K}^{\prime }$ and $\gamma _{K}$ denote respectively the
smaller coordinate angles between the un-oriented direction of the upper
trajectory photon and the direction towards the lens and between the
un-oriented direction of the lower trajectory photon and the direction
towards the lens. We have 
\begin{equation}
\gamma _{K}^{\prime }:=\arctan \!\left\vert r_{Sch\ddot{u}E}^{\prime }\frac{%
\dot{\varphi}_{Sch\ddot{u}E}^{\prime }}{\dot{r}_{Sch\ddot{u}E}^{\prime }}%
\right\vert =\arctan \!\frac{\chi _{P}^{\prime }\!\left( C_{Sch\ddot{u}%
E}^{\prime }+g{}_{Sch\ddot{u}}^{\prime }\right) ^{-1}}{\cos \chi _{Sch\ddot{u%
}}\sin \chi _{Sch\ddot{u}}},  \label{smallangle}
\end{equation}%
and 
\begin{equation}
\gamma _{K}:=\arctan \!\left\vert r_{Sch\ddot{u}E}\frac{\dot{\varphi}_{Sch%
\ddot{u}E}}{\dot{r}_{Sch\ddot{u}E}}\right\vert =\arctan \!\frac{\chi
_{P}\left( C_{Sch\ddot{u}E}+g_{Sch\ddot{u}}\right) ^{-1}}{\cos \chi _{Sch%
\ddot{u}}\sin \chi _{Sch\ddot{u}}}.
\end{equation}

\subsection{Null geodesics inside the Kottler vacuole}

In this region where prevails Kottler metric, we shall not go deeper in the
details leading to the same results already discussed in the flat case \cite%
{TSch1,TD-ES}, and what we would have to do is to calculate first the scale
factors $a_{Sch\ddot{u}S}^{\prime }$ and $a_{Sch\ddot{u}S}$ and then their
corresponding times $t_{Sch\ddot{u}S}^{\prime }$ and $t_{Sch\ddot{u}S}$, at
which the two photons enter inside the vacuole, by integrating the Friedman
equation with final conditions at the exit points from the vacuole.

The partially integrated geodesic equations for the upper photon in Kottler
space-time (\ref{kottmetric}) are 
\begin{equation}
\dot{T}\,=\frac{1}{B(r)},\quad \dot{r}=\pm \left( 1-\frac{r_{P}^{\prime 2}}{%
r^{2}}\frac{B(r)}{B(r_{P}^{\prime })}\right) ^{1/2},\quad \dot{\varphi}=%
\frac{r_{P}^{\prime }}{r^{2}\sqrt{B(r_{P}^{\prime })}},  \label{sol2}
\end{equation}%
where $r_{P}^{\prime }$ is the peri-lens. The travel time of the upper
photon inside the vacuole from the entry point to the exit point can be
obtained by using the relation (\ref{match-cond}) between the Schwarzschild
time $T$ and the Friedmann time $t$, 
\begin{eqnarray}
T_{Sch\ddot{u}E}^{\prime }-T_{Sch\ddot{u}S}^{\prime } &=&\cos \chi _{Sch%
\ddot{u}}\int_{t_{Sch\ddot{u}S}^{\prime }}^{t_{Sch\ddot{u}E}^{\prime }}\frac{%
dt}{B_{Sch\ddot{u}}(t)} \\
&=&\cos \chi _{Sch\ddot{u}}\int_{a_{Sch\ddot{u}S}^{\prime }}^{a_{Sch\ddot{u}%
E}^{\prime }}\frac{da}{B_{Sch\ddot{u}}(a)f(a)},  \label{ShuS1}
\end{eqnarray}%
owing to $dt=da/f(a)$. Another expression for this travel time can be
obtained by making use of the well known equation 
\begin{equation}
dT=\pm \frac{dr}{v^{\prime }(r)},\quad v^{\prime }(r):=B(r)\sqrt{1-\frac{%
r_{P}^{\prime 2}}{r^{2}}\frac{B(r)}{B(r_{P}^{\prime })}},  \label{exp1}
\end{equation}%
which follows immediately by eliminating the affine parameter between $\dot{T%
}$ and $\dot{r}$ in (\ref{sol2}). The peri-lens $r_{P}^{\prime }$ is given
approximately by \cite{TSch1} 
\begin{equation}
r_{P}^{\prime }\simeq r_{Sch\ddot{u}E}^{\prime }\sin \gamma _{KE}^{\prime
}-M/8\pi ,  \label{perilensK1}
\end{equation}%
which is obtained by developing the smaller coordinate angle $\gamma
_{K}^{\prime }$ (\ref{smallangle}), using $\dot{r}$ and $\dot{\varphi}$ of (%
\ref{sol2}), to first order in the ratio of Schwarzschild radius to
peri-lens $M/4\pi r_{P}^{\prime }$, which in our case is of order $10^{-5}$.
The integral of the right-hand side of (\ref{exp1}) can be split in two
integrals according to the fact that $r$ decreases with time from $r_{Sch%
\ddot{u}S}^{\prime }$ to $r_{P}^{\prime }$ while it increases from $%
r_{P}^{\prime }$ to $r_{Sch\ddot{u}E}^{\prime }$, i.e. 
\begin{equation}
T_{Sch\ddot{u}E}^{\prime }-T_{Sch\ddot{u}S}^{\prime }=\int_{r_{P}^{\prime
}}^{r_{Sch\ddot{u}E}^{\prime }}\frac{dr}{v^{\prime }(r)}+\int_{r_{P}^{\prime
}}^{r_{Sch\ddot{u}S}^{\prime }}\frac{dr}{v^{\prime }(r)},  \label{ShuS2}
\end{equation}%
where $r_{Sch\ddot{u}E}^{\prime }$ and $r_{Sch\ddot{u}S}^{\prime }$ are
related respectively to $a_{Sch\ddot{u}E}^{\prime }$ and $a_{Sch\ddot{u}%
S}^{\prime }$ by the matching condition (\ref{matchcond}). It follows from (%
\ref{ShuS1}) and (\ref{ShuS2}) 
\begin{equation}
\int_{r_{P}^{\prime }}^{a_{Sch\ddot{u}E}^{\prime }\sin \chi _{Sch\ddot{u}}}%
\frac{dr}{v^{\prime }(r)}+\int_{r_{P}^{\prime }}^{a_{Sch\ddot{u}S}^{\prime
}\sin \chi _{Sch\ddot{u}}}\frac{dr}{v^{\prime }(r)}=\cos \chi _{Sch\ddot{u}%
}\int_{a_{Sch\ddot{u}S}^{\prime }}^{a_{Sch\ddot{u}E}^{\prime }}\frac{da}{%
B_{Sch\ddot{u}}(a)f(a)},
\end{equation}%
which enables us to obtain $a_{Sch\ddot{u}S}^{\prime }$ by numerical
integration. Then we have to integrate numerically the Friedmann equation to
obtain $t_{Sch\ddot{u}S}^{\prime }$, i.e. 
\begin{equation}
t_{Sch\ddot{u}S}^{\prime }=\int_{a_{0}}^{a_{Sch\ddot{u}S}^{\prime }}\dfrac{da%
}{f(a)}.
\end{equation}

Following the same reasoning as for the upper trajectory, we get for the
lower trajectory similar formulae, i.e. 
\begin{equation}
t_{Sch\ddot{u}S}=\int_{a_{0}}^{a_{Sch\ddot{u}S}}\dfrac{da}{f(a)},
\end{equation}%
with $a_{Sch\ddot{u}S}$ calculated by solving numerically the equation 
\begin{equation}
\int_{r_{P}}^{a_{Sch\ddot{u}E}\sin \chi _{Sch\ddot{u}}}\frac{dr}{v(r)}%
+\int_{r_{P}}^{a_{Sch\ddot{u}S}\sin \chi _{Sch\ddot{u}}}\frac{dr}{v(r)}=\cos
\chi _{Sch\ddot{u}}\int_{a_{Sch\ddot{u}S}}^{a_{Sch\ddot{u}E}}\frac{da}{B_{Sch%
\ddot{u}}(a)f(a)},
\end{equation}%
where $r_{P}$ is also expressed by a formula similar to (\ref{perilensK1}), 
\begin{equation}
r_{P}\simeq r_{Sch\ddot{u}E}\sin \gamma _{KE}-M/8\pi .
\end{equation}

We can avoid computing numerically the time $t_{Sch\ddot{u}S}$ for lower
trajectory photon following the same approximation method described in \cite%
{TD-ES}. It could be determined by difference with $t_{Sch\ddot{u}S}^{\prime
}$ via an approximate analytical expression. First, thanks to the relation (%
\ref{match-cond}) between the Schwarzschild time $T$ and the Friedmann time $%
t$, we may express the difference in the travel times of both photons inside
the vacuole as 
\begin{eqnarray}
T_{Sch\ddot{u}E}^{\prime }-T_{Sch\ddot{u}S}^{\prime }-(T_{Sch\ddot{u}%
E}-T_{Sch\ddot{u}S}) &=&\cos \chi _{Sch\ddot{u}}\int_{t_{Sch\ddot{u}%
E}}^{t_{Sch\ddot{u}E}^{\prime }}\frac{dt}{B_{Sch\ddot{u}}(t)}  \nonumber \\
&&-\cos \chi _{Sch\ddot{u}}\int_{t_{Sch\ddot{u}S}}^{t_{Sch\ddot{u}S}^{\prime
}}\frac{dt}{B_{Sch\ddot{u}}(t)} \\
&\simeq &\cos \chi _{Sch\ddot{u}}\frac{t_{Sch\ddot{u}E}^{\prime }-t_{Sch%
\ddot{u}E}}{B_{Sch\ddot{u}E}^{\prime }}  \nonumber \\
&&-\cos \chi _{Sch\ddot{u}}\frac{t_{Sch\ddot{u}S}^{\prime }-t_{Sch\ddot{u}S}%
}{B_{Sch\ddot{u}S}^{\prime }},  \label{diftrav1}
\end{eqnarray}%
where we have used the fact that $B_{Sch\ddot{u}}$ vary appreciably only on
cosmological time intervals with $B_{Sch\ddot{u}E}^{\prime }:=B_{Sch\ddot{u}%
}(t_{Sch\ddot{u}E}^{\prime })$ and $B_{Sch\ddot{u}E}^{\prime }:=B_{Sch\ddot{u%
}}(t_{Sch\ddot{u}E}^{\prime })$. Secondly, we can make use of (\ref{exp1})
to write this difference in travel times as 
\begin{equation}
T_{Sch\ddot{u}E}^{\prime }-T_{Sch\ddot{u}S}^{\prime }-(T_{Sch\ddot{u}%
E}-T_{Sch\ddot{u}S})=\Delta T_{K}+\int_{r_{Sch\ddot{u}E}}^{r_{Sch\ddot{u}%
E}^{\prime }}\frac{dr}{v(r)}+\int_{r_{Sch\ddot{u}S}}^{r_{Sch\ddot{u}%
S}^{\prime }}\frac{dr}{v(r)},  \label{diftrav2}
\end{equation}%
where we have broken up the integrals to produce the following expression 
\begin{equation}
\Delta T_{K}=\int_{r_{P}^{\prime }}^{r_{Sch\ddot{u}E}^{\prime }}\frac{dr}{%
v^{\prime }(r)}+\int_{r_{P}^{\prime }}^{r_{Sch\ddot{u}S}^{\prime }}\frac{dr}{%
v^{\prime }(r)}-\left( \int_{r_{P}}^{r_{Sch\ddot{u}E}^{\prime }}\frac{dr}{%
v(r)}+\int_{r_{P}}^{r_{Sch\ddot{u}S}^{\prime }}\frac{dr}{v(r)}\right) .
\label{TK}
\end{equation}%
We may interpret this as the difference in the travel times between the
upper photon and an imaginary lower photon that starts from the same point
as the upper photon $r_{Sch\ddot{u}S}^{\prime }$, deflected by the lens at
peri-lens $r_{P}$ and finally arrives at the same point as the upper photon $%
r_{Sch\ddot{u}E}^{\prime }$. This latter (\ref{TK}) is almost identical to
an expression already involved in the calculation of time delay in Kottler
solution \cite{KTD}, just replace $r_{T}$ by $r_{Sch\ddot{u}E}^{\prime }$, $%
r_{S}$ by $r_{Sch\ddot{u}S}^{\prime }$, $r_{0}^{\prime }$ by $r_{P}^{\prime
} $ and $r_{0}$ by $r_{P}$. The result is 
\begin{eqnarray}
\Delta T_{K} &\simeq &\frac{r_{P}^{2}-r_{P}^{\prime 2}}{2}\left( \dfrac{1}{%
r_{Sch\ddot{u}E}^{\prime }}+\dfrac{1}{r_{Sch\ddot{u}S}^{\prime }}\right)
\!+\!\frac{M}{2\pi }\ln \dfrac{r_{P}}{r_{P}^{\prime }}-3\left( \frac{M}{4\pi
r_{P}^{\prime }}\right) ^{2}\!\frac{r_{P}^{2}-r_{P}^{\prime 2}}{8r_{P}^{2}}%
\times  \nonumber \\
&&\sqrt{\frac{3}{\Lambda }}\left[ \func{arctanh}\left( \sqrt{\frac{\Lambda }{%
3}}r_{Sch\ddot{u}E}^{\prime }\right) +\func{arctanh}\left( \sqrt{\frac{%
\Lambda }{3}}r_{Sch\ddot{u}S}^{\prime }\right) \right] .  \label{KTD}
\end{eqnarray}%
Furthermore, since we deal with smaller length and time scales than
cosmological ones, 
\begin{equation}
\int_{r_{Sch\ddot{u}E}}^{r_{Sch\ddot{u}E}^{\prime }}\frac{dr}{v(r)}\simeq 
\frac{r_{Sch\ddot{u}E}^{\prime }-r_{Sch\ddot{u}E}}{v^{\prime }{}_{Sch\ddot{u}%
E}},\quad \int_{r_{Sch\ddot{u}S}}^{r_{Sch\ddot{u}S}^{\prime }}\frac{dr}{%
v^{\prime }(r)}\simeq \frac{r_{Sch\ddot{u}S}^{\prime }-r_{Sch\ddot{u}S}}{%
v{}_{Sch\ddot{u}S}^{\prime }},  \label{smalllength}
\end{equation}%
and, using the Friedmann equation, 
\begin{eqnarray}
r_{Sch\ddot{u}E}^{\prime }-r_{Sch\ddot{u}E} &=&(a_{Sch\ddot{u}E}^{\prime
}-a_{Sch\ddot{u}E})\sin \chi _{Sch\ddot{u}}  \nonumber \\
&\simeq &f_{Sch\ddot{u}E}^{\prime }\sin \chi _{Sch\ddot{u}}(t_{Sch\ddot{u}%
E}^{\prime }-t_{Sch\ddot{u}E}), \\
r_{Sch\ddot{u}S}^{\prime }-r_{Sch\ddot{u}S} &=&(a_{Sch\ddot{u}S}^{\prime
}-a_{Sch\ddot{u}S})\sin \chi _{Sch\ddot{u}}  \nonumber \\
&\simeq &f_{Sch\ddot{u}S}^{\prime }\sin \chi _{Sch\ddot{u}}(t_{Sch\ddot{u}%
S}^{\prime }-t_{Sch\ddot{u}S}).
\end{eqnarray}%
with $v^{\prime }{}_{Sch\ddot{u}E}:=v(r_{Sch\ddot{u}E}^{\prime })$, $v{}_{Sch%
\ddot{u}S}^{\prime }:=v(r_{Sch\ddot{u}S}^{\prime })$, $f_{Sch\ddot{u}%
E}^{\prime }:=f(a_{Sch\ddot{u}E}^{\prime })$ and $f_{Sch\ddot{u}S}^{\prime
}:=f(a_{Sch\ddot{u}S}^{\prime })$. Using this together with (\ref{KTD}), (%
\ref{diftrav2}) and (\ref{diftrav1}), we therefore get 
\begin{equation}
t_{Sch\ddot{u}S}-t_{Sch\ddot{u}S}^{\prime }\simeq \frac{\Delta T_{K}+\left( 
\dfrac{f_{Sch\ddot{u}E}^{\prime }\sin \chi _{Sch\ddot{u}}}{v^{\prime }{}_{Sch%
\ddot{u}E}}-\dfrac{\cos \chi _{Sch\ddot{u}}}{B_{Sch\ddot{u}E}^{\prime }}%
\right) (t_{Sch\ddot{u}E}^{\prime }-t_{Sch\ddot{u}E})}{\left( \dfrac{f_{Sch%
\ddot{u}S}^{\prime }\sin \chi _{Sch\ddot{u}}}{v{}_{Sch\ddot{u}S}^{\prime }}+%
\dfrac{\cos \chi _{Sch\ddot{u}}}{B_{Sch\ddot{u}S}^{\prime }}\right) }.
\end{equation}%
Then the knowledge of $t_{Sch\ddot{u}S}^{\prime }$ allows one to deduce $%
t_{Sch\ddot{u}S}$ where $t_{Sch\ddot{u}E}^{\prime }-t_{Sch\ddot{u}E}$ is
given by (\ref{difexit}). It should be noted that the lower photon enters
the vacuole after the upper one, even though it leaves the vacoule before.
Admittedly, this is caused by the fact that the upper photon undergoes more
strongly the gravitational effect since it passes closest to the lens ($%
r_{P}^{\prime }<r_{P}$) as shown in Fig. \ref{def}.

We should also compute the angles $\varphi _{Sch\ddot{u}S}^{\prime }$ and $%
\varphi _{Sch\ddot{u}S}$ as well as the 4-velocities at points of entry into
the vacuole, needed for the next section. It would be necessary to make use
of the well known equation 
\begin{equation}
d\varphi =\pm \frac{dr}{u^{\prime }(r)},\quad u^{\prime }(r):=r\sqrt{\frac{%
r^{2}}{r_{P}^{\prime 2}}-1}\sqrt{1-\dfrac{M}{4\pi r_{P}^{\prime }}\left( 
\dfrac{r_{P}^{\prime }}{r}+\frac{r}{r+r_{P}^{\prime }}\right) },
\label{exp2}
\end{equation}%
which follows from (\ref{sol2}), where the cosmological constant $\Lambda $
incidentally disappeared. Due to the fact that the angle $\varphi $
increases when the upper photon approaches the lens as well as when it moves
away, the integral of the right-hand side of (\ref{exp2}) along the
trajectory may thus be split up as follows, 
\begin{equation}
\varphi _{Sch\ddot{u}E}^{\prime }-\varphi _{Sch\ddot{u}S}^{\prime
}=\int_{r_{P}^{\prime }}^{r_{Sch\ddot{u}E}^{\prime }}\frac{dr}{u^{\prime }(r)%
}+\int_{r_{P}^{\prime }}^{r_{Sch\ddot{u}S}^{\prime }}\frac{dr}{u^{\prime }(r)%
},
\end{equation}%
and gives, to first order in the ratio $M/4\pi r_{P}^{\prime }$, 
\begin{eqnarray}
\varphi _{Sch\ddot{u}S}^{\prime } &\simeq &\!\varphi _{Sch\ddot{u}E}^{\prime
}-\pi +\arcsin \dfrac{r_{P}^{\prime }}{r_{Sch\ddot{u}E}^{\prime }}+\arcsin 
\dfrac{r_{P}^{\prime }}{r_{Sch\ddot{u}S}^{\prime }}-\dfrac{M}{8\pi
r_{P}^{\prime }}\left( \sqrt{1\!-\!\dfrac{r_{P}^{\prime 2}}{r_{Sch\ddot{u}%
E}^{\prime 2}}}+\right.  \nonumber \\
&&\left. \!\sqrt{1\!-\!\dfrac{r_{P}^{\prime 2}}{r_{Sch\ddot{u}S}^{\prime 2}}}%
+\sqrt{\dfrac{r_{Sch\ddot{u}E}^{\prime }-\!r_{P}^{\prime }}{r_{Sch\ddot{u}%
E}^{\prime }+\!r_{P}^{\prime }}}+\sqrt{\dfrac{r_{Sch\ddot{u}S}^{\prime
}-\!r_{P}^{\prime }}{r_{Sch\ddot{u}S}^{\prime }+\!r_{P}^{\prime }}}\right)
\!\,,
\end{eqnarray}%
where $\varphi _{Sch\ddot{u}E}^{\prime }$ is given by (\ref{varphiup}). In
the same manner, we obtain for the lower trajectory, 
\begin{eqnarray}
\varphi _{Sch\ddot{u}S} &\simeq &\!\varphi _{Sch\ddot{u}E}+\pi -\arcsin 
\dfrac{r_{P}}{r_{Sch\ddot{u}E}}-\arcsin \dfrac{r_{P}}{r_{Sch\ddot{u}S}}+%
\dfrac{M}{8\pi r_{P}}\left( \sqrt{1\!-\!\dfrac{r_{P}^{2}}{r_{Sch\ddot{u}%
E}^{2}}}+\right.  \nonumber \\
&&\left. \!\sqrt{1\!-\!\dfrac{r_{P}^{2}}{r_{Sch\ddot{u}S}^{2}}}+\sqrt{\dfrac{%
r_{Sch\ddot{u}E}-\!r_{P}}{r_{Sch\ddot{u}E}+\!r_{P}}}+\sqrt{\dfrac{r_{Sch%
\ddot{u}S}-\!r_{P}}{r_{Sch\ddot{u}S}+\!r_{P}}}\right) \!\,,
\label{varphilos}
\end{eqnarray}%
where $\varphi _{Sch\ddot{u}E}$ is given by (\ref{varphilow}). On account of
(\ref{sol2}), the two 4-velocities of the upper and lower photons at points
of entry into the vacuole are respectively, 
\begin{equation}
\dot{T}\,_{Sch\ddot{u}S}^{\prime }=\frac{1}{B_{Sch\ddot{u}S}^{\prime }}%
,\quad \dot{r}_{Sch\ddot{u}S}^{\prime }=-\sqrt{1-\dfrac{r_{P}^{\prime
2}B_{Sch\ddot{u}S}^{\prime }}{r_{Sch\ddot{u}S}^{\prime 2}B(r_{P}^{\prime })}}%
,\quad \dot{\varphi}_{Sch\ddot{u}S}^{\prime }=\!\!\,\dfrac{r_{P}^{\prime }}{%
r_{Sch\ddot{u}S}^{\prime 2}\sqrt{B(r_{P}^{\prime })}^{\!}},
\end{equation}%
and 
\begin{equation}
\dot{T}\,_{Sch\ddot{u}S}=\frac{1}{B_{Sch\ddot{u}S}},\quad \dot{r}_{Sch\ddot{u%
}S}=-\sqrt{1-\dfrac{r_{P}^{2}B_{Sch\ddot{u}S}}{r_{Sch\ddot{u}S}^{2}\,B(r_{P})%
}},\quad \dot{\varphi}_{Sch\ddot{u}S}=\dfrac{-r_{P}}{r_{Sch\ddot{u}S}^{2}%
\sqrt{B(r_{P})}^{\!}},
\end{equation}%
in Schwarzschild coordinates, where we have taken into account that $r$ is
decreasing with time since both photons approach the lens, while $\varphi $
is increasing with time for the upper photon and decreasing with time for
the lower photon. These two 4-velocities can be translated, thanks to the
Jacobian (\ref{Jacob}), into Friedmann coordinates, 
\begin{eqnarray}
&&\dot{t}_{Sch\ddot{u}S}^{\prime }=\frac{\cos \chi _{Sch\ddot{u}}}{B_{Sch%
\ddot{u}S}^{\prime }}\left( 1+C_{Sch\ddot{u}S}^{\prime }\sqrt{1-\dfrac{%
r_{P}^{\prime 2}B_{Sch\ddot{u}S}^{\prime }}{r_{Sch\ddot{u}S}^{\prime
2}B(r_{P}^{\prime })}}\right) , \\
&&\dot{\chi}_{Sch\ddot{u}S}^{\prime }=\dfrac{-\cos \chi _{Sch\ddot{u}}}{%
a_{Sch\ddot{u}S}^{\prime }B_{Sch\ddot{u}S}^{\prime }}\left( C_{Sch\ddot{u}%
S}^{\prime }+\sqrt{1-\dfrac{r_{P}^{\prime 2}B_{Sch\ddot{u}S}^{\prime }}{%
r_{Sch\ddot{u}S}^{\prime 2}B(r_{P}^{\prime })}}\right) ,
\end{eqnarray}%
for the upper trajectory and, 
\begin{eqnarray}
&&\dot{t}_{Sch\ddot{u}S}=\frac{\cos \chi _{Sch\ddot{u}}}{B_{Sch\ddot{u}S}}%
\left( 1+C_{Sch\ddot{u}S}\sqrt{1-\dfrac{r_{P}^{2}B_{Sch\ddot{u}S}}{r_{Sch%
\ddot{u}S}^{2}\,B(r_{P})}}\right) , \\
&&\dot{\chi}_{Sch\ddot{u}S}=\dfrac{-\cos \chi _{Sch\ddot{u}}}{a_{Sch\ddot{u}%
S}B_{Sch\ddot{u}S}}\left( C_{Sch\ddot{u}S}+\sqrt{1-\dfrac{r_{P}^{2}B_{Sch%
\ddot{u}S}}{r_{Sch\ddot{u}S}^{2}\,B(r_{P})}}\right) ,
\end{eqnarray}%
for the lower trajectory.

\subsection{Null geodesics between the source and the Sch\"{u}cking sphere}

Again, we should follow in this region where prevails Friedmann metric, the
same procedure as previously in the subsection~\ref{1}. The integration of
the two geodesic equations for the upper photon, (\ref{feqt1}) and (\ref%
{feqt3}), with final conditions at the point of entry into the vacuole, 
\begin{equation}
\begin{array}{lll}
t=t_{Sch\ddot{u}S}^{\prime }, & \chi =\chi _{Sch\ddot{u}}, & \varphi
=\varphi _{Sch\ddot{u}S}^{\prime }, \\ 
\dot{t}=\dot{t}_{Sch\ddot{u}S}^{\prime }, & \dot{\chi}=\dot{\chi}_{Sch\ddot{u%
}S}^{\prime } & \dot{\varphi}=\dot{\varphi}_{Sch\ddot{u}S}^{\prime },%
\end{array}%
\end{equation}%
gives 
\begin{equation}
\dot{t}=\frac{E^{\prime }}{a(t)},\quad \dot{\varphi}\!=\dfrac{J^{\prime }}{%
a^{2}\sin ^{2}\chi },\quad \varphi =\varphi _{Sch\ddot{u}S}^{\prime
}+\arcsin \dfrac{(J^{\prime }/E^{\prime })\cot \chi }{\sqrt{1-(J^{\prime
}/E^{\prime })^{2}}}-\gamma ^{\prime },  \label{sol3}
\end{equation}%
where the constants $E^{\prime }$, $J^{\prime }$ and $\gamma ^{\prime }$ are
defined by%
\begin{equation}
\begin{array}{l}
E^{\prime }:=\dfrac{a_{Sch\ddot{u}S}^{\prime }\cos \chi _{Sch\ddot{u}}}{%
B_{Sch\ddot{u}S}^{\prime }}\left( 1+C_{Sch\ddot{u}S}^{\prime }\sqrt{1-\dfrac{%
r_{P}^{\prime 2}B_{Sch\ddot{u}S}^{\prime }}{r_{Sch\ddot{u}S}^{\prime
2}B(r_{P}^{\prime })}}\right) , \\ 
J^{\prime }:=\dfrac{r_{P}^{\prime }}{\sqrt{B(r_{P}^{\prime })}},\quad \gamma
^{\prime }:=\arcsin \dfrac{(J^{\prime }/E^{\prime })\cot \chi _{Sch\ddot{u}}%
}{\sqrt{1-(J^{\prime }/E^{\prime })^{2}}}.%
\end{array}%
\end{equation}%
We then deduce the expression of the angle $\varphi _{S}^{\prime }$, 
\begin{equation}
\varphi _{S}^{\prime }=\varphi _{Sch\ddot{u}S}^{\prime }+\arcsin \dfrac{%
(J^{\prime }/E^{\prime })\cot \chi _{L,S}}{\sqrt{1-(J^{\prime }/E^{\prime
})^{2}}}-\gamma ^{\prime },  \label{varpsup}
\end{equation}%
where the geodesic distance $\chi _{L,S}$ between the lens and the source is
approximated by 
\[
\chi _{L,S}\simeq \chi _{S}-\chi _{L}, 
\]%
owing to the fact that the deflection angle $-\varphi _{S}$ is of the order
of a few arcseconds ($\sim 10^{-5}$). For the lower photon, we just have to
replace $J^{\prime }$ by $-J$ and then obtain the solution of the geodesic
equations. We only give here the expression of the angle $\varphi _{S}$, 
\begin{equation}
\varphi _{S}=\varphi _{Sch\ddot{u}S}-\arcsin \dfrac{(J/E)\cot \chi _{L,S}}{%
\sqrt{1-(J/E)^{2}}}+\gamma ,  \label{varpslo}
\end{equation}%
where $\varphi _{Sch\ddot{u}S}$ is given by (\ref{varphilos}) and 
\begin{equation}
\begin{array}{l}
E:=\dfrac{a_{Sch\ddot{u}S}\cos \chi _{Sch\ddot{u}}}{B_{Sch\ddot{u}S}}\left(
1+C_{Sch\ddot{u}S}\sqrt{1-\dfrac{r_{P}^{2}B_{Sch\ddot{u}S}}{r_{Sch\ddot{u}%
S}^{2}B(r_{P})}}\right) , \\ 
J:=\dfrac{r_{P}}{\sqrt{B(r_{P})}},\quad \gamma :=\arcsin \dfrac{(J/E)\cot
\chi _{Sch\ddot{u}}}{\sqrt{1-(J/E)^{2}}}.%
\end{array}%
\end{equation}

The angles $\varphi _{S}^{\prime }$ and $\varphi _{S}$ must be equal due to
the fact that both photons are emitted by the same source. Among all the
parameters involved in the calculation of these angles, their equality could
only be ensured by a fixed value of the mass, i.e. we have to vary $M$ in
order to get $\varphi _{S}=\varphi _{S}^{\prime }$. Once the correct mass is
determined, one should be able to compute the time delay between both
photons $t_{S}-t_{S}^{\prime }$ (difference between their total travel
times). To achieve this, we first need to compute the distance traveled by
the upper photon between the source and its entry point into the vacuole ($%
\chi _{S,Sch\ddot{u}S}^{\prime }$). Differentiating the solution $\varphi $ (%
\ref{sol3}) with respect to $\chi $ and substituting into the closed
Friedmann metric (\ref{closed-fried-metric}), one easily get an equation
similar to (\ref{varpchi}), by putting $J^{\prime }/E^{\prime }$ instead of $%
\chi _{P}^{\prime }$, 
\begin{equation}
\int \frac{dt}{a(t)}=-\int \frac{d\chi }{h^{\prime }(\chi )},\quad h^{\prime
}(\chi )=\sqrt{1-(J^{\prime }/E^{\prime })^{2}/\sin ^{2}\chi },
\end{equation}%
where the negative sign indicates that $\chi $ between the source and the
vacuole is decreasing with time. Obviously, The evaluation of the right-hand
side of the last equation between $\chi _{L,S}$ and $\chi _{Sch\ddot{u}}$
gives the geodesic distance $\chi _{S,Sch\ddot{u}S}^{\prime }$, 
\begin{equation}
\int_{a_{S}^{\prime }}^{a_{Sch\ddot{u}S}^{\prime }}\frac{da}{af(a)}=\arcsin 
\frac{\cos \chi _{Sch\ddot{u}}}{\sqrt{1-(J^{\prime }/E^{\prime })^{2}}}%
-\arcsin \frac{\cos \chi _{L,S}}{\sqrt{1-(J^{\prime }/E^{\prime })^{2}}},
\label{form4}
\end{equation}%
where we have inserted the function $f(a)$ in the left-hand side via $%
dt=da/f(a)$. Solving numerically this equation, one can deduce $a_{Sch\ddot{u%
}S}^{\prime }$. Then, if one is interested in $t_{S}^{\prime }$, it suffices
to use the Friedmann equation, i.e. 
\begin{equation}
t_{S}^{\prime }=\int_{a_{0}}^{a_{S}^{\prime }}\dfrac{da}{f(a)}.
\end{equation}%
Likewise, the emission time $t_{S}$ of the lower photon is obtained by
numerical integration of Friedmann equation, i.e. 
\begin{equation}
t_{S}=\int_{a_{0}}^{a_{S}}\dfrac{da}{f(a)},
\end{equation}%
where $a_{S}$ is calculated by solving numerically the equation 
\begin{equation}
\int_{a_{S}}^{a_{Sch\ddot{u}S}}\frac{da}{af(a)}=\arcsin \frac{\cos \chi _{Sch%
\ddot{u}}}{\sqrt{1-(J/E)^{2}}}-\arcsin \frac{\cos \chi _{L,S}}{\sqrt{%
1-(J/E)^{2}}}.  \label{form5}
\end{equation}

However, one can proceed, as in the flat case \cite{TD-ES}, in a different
manner computing directly the time delay through an approximate analytical
expression. Combining (\ref{form4}) and (\ref{form5}), one gets, after
evaluating the integrals with respect to time, 
\begin{eqnarray}
\int_{t_{Sch\ddot{u}S}^{\prime }}^{t_{Sch\ddot{u}S}}\frac{dt}{a(t)}%
-\int_{t_{S}^{\prime }}^{t_{S}}\frac{dt}{a(t)} &=&\arcsin \frac{\cos \chi
_{Sch\ddot{u}}}{\sqrt{1-(J/E)^{2}}}-\arcsin \frac{\cos \chi _{L,S}}{\sqrt{%
1-(J/E)^{2}}}-  \nonumber \\
&&\arcsin \frac{\cos \chi _{Sch\ddot{u}}}{\sqrt{1-(J^{\prime }/E^{\prime
})^{2}}}+\arcsin \frac{\cos \chi _{L,S}}{\sqrt{1-(J^{\prime }/E^{\prime
})^{2}}}.
\end{eqnarray}%
Using the fact that the scale factor $a(t)$ varies noticeably only over
cosmological time scales, the left-hand side of this equation can be
approximated by 
\begin{equation}
\text{LHS}\simeq \frac{t_{Sch\ddot{u}S}-t_{Sch\ddot{u}S}^{\prime }}{a_{Sch%
\ddot{u}S}^{\prime }}-\frac{t_{S}-t_{S}^{\prime }}{a_{S}^{\prime }}.
\end{equation}%
Expanding the right-hand side to second order in $J^{\prime }/E^{\prime }$
and $J/E$ ($\sim 10^{-6}$ in our case), on account of (\ref{varpsup}) and (%
\ref{varpslo}), one gets 
\begin{eqnarray}
\text{RHS} &\simeq &\frac{1}{2}(\cot \chi _{Sch\ddot{u}}-\cot \chi _{L,S})%
\left[ \left( \frac{J}{E}\right) ^{2}-\left( \frac{J^{\prime }}{E^{\prime }}%
\right) ^{2}\right]  \\
&\simeq &\frac{1}{2}\frac{(\varphi _{Sch\ddot{u}S}-\varphi
_{S})^{2}-(\varphi _{Sch\ddot{u}S}^{\prime }-\varphi _{S})^{2}}{\cot \chi
_{Sch\ddot{u}}-\cot \chi _{L,S}},
\end{eqnarray}%
where $\varphi _{Sch\ddot{u}S}^{\prime }-\varphi _{S}$ and $|\varphi _{Sch%
\ddot{u}S}-\varphi _{S}|$ are of order $10^{-2}$. Equating the two sides,
one arrives finally at 
\begin{equation}
\Delta t:=t_{S}-t_{S}^{\prime }\simeq a_{S}^{\prime }\left[ \frac{t_{Sch%
\ddot{u}S}-t_{Sch\ddot{u}S}^{\prime }}{a_{Sch\ddot{u}S}^{\prime }}-\frac{1}{2%
}\frac{(\varphi _{Sch\ddot{u}S}-\varphi _{S})^{2}-(\varphi _{Sch\ddot{u}%
S}^{\prime }-\varphi _{S})^{2}}{\cot \chi _{Sch\ddot{u}}-\cot \chi _{L,S}}%
\right] .
\end{equation}
To sum up, the upper photon, after being the first emitted by the source,
also penetrates the first the vacuole but it leaves it the last in such a
way that it reaches the Earth at the same time with the lower photon.

\section{Application to the lensed quasar SDSS J1004+4112}

In this last step, we apply our results to the lensed quasar SDSS J1004+4112 
\cite{inada,oguri,ota,fohl,kawano} with 
\begin{equation}
\alpha ^{\prime }=5^{\prime \prime }\pm 10\%,\quad \alpha =10^{\prime \prime
}\pm 10\%,\quad z_{L}=0.68,\quad z_{S}=1.734.
\end{equation}%
The Earth-Lens and Earth-Source distances are calculated by (\ref{distES}); $%
\chi _{L}:=\chi (z_{L})$ and $\chi _{S}:=\chi (z_{S})$, with $\Lambda
=0.77\cdot 3$ $\unit{am}{}^{-2}\pm 20\%$. As we said before, the mass of the
cluster of galaxies $M$ is varied until the equality $\varphi _{S}=\varphi
_{S}^{\prime }$ is satisfied. It is worth noting that, neither a value of $%
a_{0}$ nor that of the cosmological constant $\Lambda $ could make $\varphi
_{S}$s coincide. The special case $a_{0}=5\unit{am}$ is addressed in detail
for maximum `$+$', central `$\pm 0$' and minimum values of $\alpha ^{\prime }
$, $\alpha $ and $\Lambda $. In addition, we have treated the case without
cosmological constant $\Lambda =0$. The results are recorded in Tables \ref%
{table1}, \ref{table2}, \ref{table3} and \ref{table4}. 

We can see from Tables \ref{table1}, \ref{table2} and \ref{table3} that an
increasing cosmological constant $\Lambda $ within its error bar by $20\%$
leads to a decrease of the deflection angle $-\varphi _{S}$ by about $8\%$,
but this variation only increases the cluster mass $M$ by about $1\%$. Thus,
the bending of light clearly depends on the cosmological constant. Note also
the monotonous dependence of the time delay $\Delta t$ on the cosmological
constant.

In contrast to the case of flat Einstein-Straus model \cite{TSch1,TD-ES},
where the present scale factor can be set to any value without loss of
generality, the results in Tables \ref{table5} and \ref{table6} show that
the mass, the deflection angle and time delay are somehow related to the
present scale factor parameter. They decrease significantly as the present
scale factor gets smaller and smaller below a limit value close to $7\unit{am%
}$ ($\Omega _{k0}\simeq 0.02$). This dependence on $a_{0}$ comes from the
curvature density term in Friedmann equation that vanishes in the case of
flat space ($k=0$). Nevertheless, above this limit value ($\Omega _{k0}<0.02$%
), the effect of $a_{0}$ is very small and the obtained steady values for
the mass, the deflection angle and time delay are found to be
indistinguishable from the flat case \footnote{%
For accuracy, we correct the rounding error in our previous work \cite{TD-ES}%
: the time delay, with central values of $\alpha ^{\prime }$, $\alpha $ and $%
\Lambda $, should be $9.71\unit{years}$, not $9.72\unit{years}$.}. Ditto for
all possible values of $\alpha ^{\prime }$, $\alpha $ and $\Lambda $ within
their error bars.

\begin{table}[htp]\setlength\tabcolsep{7pt}%
\caption{ Upper limit value of $\Lambda$: $\Lambda=0.77 \cdot
3\unit{am}^{-2}+ 20\%$ ($a_0=5\unit{am}$)}\label{table1}$%
\begin{tabular}{llllllllll}
\hline
$\alpha ^{\prime }\pm 10\%$ & $+$ & $+$ & $+$ & $\pm 0$ & $\pm 0$ & $\pm 0$
& $-$ & $-$ & $-$ \\ 
$\alpha \pm 10\%$ & $+$ & $\pm 0$ & $-$ & $+$ & $\pm 0$ & $-$ & $+$ & $\pm 0$
& $-$ \\ 
$M[10^{13}M_{\odot }]$ & $\mathbf{2.14}$ & $1.95$ & $1.75$ & $1.95$ & $1.77$
& $1.59$ & $1.75$ & $1.59$ & $\mathbf{1.43}$ \\ 
$-\varphi _{S}[^{\prime \prime }]$ & $9.99$ & $8.17$ & $\mathbf{6.35}$ & $%
10.89$ & $9.08$ & $7.26$ & $\mathbf{11.80}$ & $9.99$ & $8.17$ \\ 
$\Delta t[\unit{years}]$ & $11.52$ & $8.95$ & $\mathbf{6.58}$ & $12.04$ & $%
9.52$ & $7.19$ & $\mathbf{12.44}$ & $9.99$ & $7.71$ \\ \hline
\end{tabular}%
${\small \vspace{1ex}}

{\small \raggedright Fitting the cluster mass with fixed present scale
factor parameter: The angle $-\varphi _{S}$ and time delay $\Delta t$ are
calculated as functions of the central mass $M$, the present scale factor $%
a_{0}$, the angles $\alpha ^{\prime }$, $\alpha $ and of the cosmological
constant $\Lambda $. The notations `$\pm 0$', `$+$' and `$-$' stand
respectively for the central value, the upper and the lower limits. Bold
values correspond to minimal and maximal values.}

\caption{Central value of $\Lambda$: $\Lambda=0.77 \cdot3\unit{am}^{-2}$
($a_0=5\unit{am}$)}\label{table2}$%
\begin{tabular}{llllllllll}
\hline
$\alpha ^{\prime }\pm 10\%$ & $+$ & $+$ & $+$ & $\pm 0$ & $\pm 0$ & $\pm 0$
& $-$ & $-$ & $-$ \\ 
$\alpha \pm 10\%$ & $+$ & $\pm 0$ & $-$ & $+$ & $\pm 0$ & $-$ & $+$ & $\pm 0$
& $-$ \\ 
$M[10^{13}M_{\odot }]$ & $\mathbf{2.16}$ & $1.97$ & $1.77$ & $1.97$ & $1.79$
& $1.61$ & $1.77$ & $1.61$ & $\mathbf{1.45}$ \\ 
$-\varphi _{S}[^{\prime \prime }]$ & $10.95$ & $8.96$ & $\mathbf{6.97}$ & $%
11.95$ & $9.96$ & $7.97$ & $\mathbf{12.95}$ & $10.95$ & $8.96$ \\ 
$\Delta t[\unit{years}]$ & $11.47$ & $8.92$ & $\mathbf{6.56}$ & $11.98$ & $%
9.48$ & $7.17$ & $\mathbf{12.36}$ & $9.94$ & $7.68$ \\ \hline
\end{tabular}%
$

\caption{Lower limit value of $\Lambda$: $\Lambda=0.77 \cdot3\unit{am}^{-2}
- 20\%$ ($a_0=5\unit{am}$)}\label{table3}$%
\begin{tabular}{llllllllll}
\hline
$\alpha ^{\prime }\pm 10\%$ & $+$ & $+$ & $+$ & $\pm 0$ & $\pm 0$ & $\pm 0$
& $-$ & $-$ & $-$ \\ 
$\alpha \pm 10\%$ & $+$ & $\pm 0$ & $-$ & $+$ & $\pm 0$ & $-$ & $+$ & $\pm 0$
& $-$ \\ 
$M[10^{13}M_{\odot }]$ & $\mathbf{2.14}$ & $1.95$ & $1.75$ & $1.95$ & $1.77$
& $1.59$ & $1.75$ & $1.59$ & $\mathbf{1.43}$ \\ 
$-\varphi _{S}[^{\prime \prime }]$ & $11.59$ & $9.48$ & $\mathbf{7.38}$ & $%
12.64$ & $10.54$ & $8.43$ & $\mathbf{13.70}$ & $11.59$ & $9.48$ \\ 
$\Delta t[\unit{years}]$ & $11.27$ & $8.77$ & $\mathbf{6.46}$ & $11.76$ & $%
9.32$ & $7.05$ & $\mathbf{12.13}$ & $9.75$ & $7.55$ \\ \hline
\end{tabular}%
$

\caption{$\Lambda =0$ ($a_0=5\unit{am}$)}\label{table4}$%
\begin{tabular}{llllllllll}
\hline
$\alpha ^{\prime }\pm 10\%$ & $+$ & $+$ & $+$ & $\pm 0$ & $\pm 0$ & $\pm 0$
& $-$ & $-$ & $-$ \\ 
$\alpha \pm 10\%$ & $+$ & $\pm 0$ & $-$ & $+$ & $\pm 0$ & $-$ & $+$ & $\pm 0$
& $-$ \\ 
$M[10^{13}M_{\odot }]$ & $\mathbf{2.00}$ & $1.82$ & $1.64$ & $1.82$ & $1.65$
& $1.49$ & $1.64$ & $1.49$ & $\mathbf{1.34}$ \\ 
$-\varphi _{S}[^{\prime \prime }]$ & $12.99$ & $10.63$ & $\mathbf{8.27}$ & $%
14.18$ & $11.81$ & $9.45$ & $\mathbf{15.36}$ & $12.99$ & $10.63$ \\ 
$\Delta t[\unit{years}]$ & $10.36$ & $8.07$ & $\mathbf{5.95}$ & $10.80$ & $%
8.57$ & $6.49$ & $\mathbf{11.12}$ & $8.96$ & $6.94$ \\ \hline
\end{tabular}%
$%
\end{table}%

\begin{table}[htp]\setlength\tabcolsep{7pt}%
\caption{$\alpha^{\prime}=5^{\prime\prime}$,
$\alpha=10^{\prime\prime}$, $\Lambda=0.77\cdot3\unit{am}^{-2}$}\label{table5}%
$%
\begin{tabular}{lllllllllll}
\hline
$a_{0}[\unit{am}]$ & $0.1$ & $0.3$ & $0.5$ & $0.7$ & $1$ & $2$ & $3$ & $4$ & 
$5$ & $6$ \\ 
$M\,[10^{13}M_{\odot }]$ & $0.04$ & $0.29$ & $0.62$ & $0.91$ & $1.22$ & $1.62
$ & $1.73$ & $1.77$ & $1.79$ & $1.80$ \\ 
$-\varphi _{S}[^{\prime \prime }]$ & $4.03$ & $6.64$ & $8.05$ & $8.82$ & $%
9.38$ & $9.85$ & $9.93$ & $9.95$ & $9.96$ & $9.96$ \\ 
$\Delta t[\unit{years}]$ & $0.17$ & $1.30$ & $2.90$ & $4.40$ & $6.08$ & $8.44
$ & $9.10$ & $9.36$ & $9.48$ & $9.55$ \\ \hline
\end{tabular}%
${\small \vspace{1ex}}

{\small \raggedright Fitting the cluster mass with varied present scale
factor parameter: Variation of the central mass $M$, the angle $-\varphi _{S}
$ and time delay $\Delta t$ according to different values of the present
scale factor $a_{0}$ under the above central values of the parameters} $%
\alpha ^{\prime }$, $\alpha $ and $\Lambda ${\small .}

\caption{$\alpha^{\prime}=5^{\prime\prime}$,
$\alpha=10^{\prime\prime}$, $\Lambda=0.77\cdot3\unit{am}^{-2}$}\label{table6}%
$%
\begin{tabular}{lllllllllll}
\hline
$a_{0}[\unit{am}]$ & $7$ & $8$ & $9$ & $10$ & $11$ & $12$ & $15$ & $19$ & $30
$ & $100$ \\ 
$M\,[10^{13}M_{\odot }]$ & $1.80$ & $1.81$ & $1.81$ & $1.81$ & $1.81$ & $1.82
$ & $1.82$ & $1.82$ & $1.82$ & $1.82$ \\ 
$-\varphi _{S}[^{\prime \prime }]$ & $9.97$ & $9.97$ & $9.97$ & $9.97$ & $%
9.97$ & $9.97$ & $9.97$ & $9.97$ & $9.97$ & $9.97$ \\ 
$\Delta t[\unit{years}]$ & $9.59$ & $9.62$ & $9.64$ & $9.65$ & $9.66$ & $9.67
$ & $9.69$ & $9.70$ & $9.71$ & $9.71$ \\ \hline
\end{tabular}%
$%
\end{table}%

\section{Conclusion}

In this paper, we have computed the bending of light and time delay in the
positively curved Einstein-Straus model with a positive cosmological
constant. Unlike the case of flat Einstein-Straus model in which the results
remain the same regardless of which value is chosen for the present scale
factor \cite{TSch1,TD-ES}, we show that the bending of light and time delay
decrease considerably as the present scale factor becomes smaller than a
limit value about $7\unit{am}$. The same holds true for the mass of the
lens. But, for larger values of the present scale factor, the results are
almost compatible with those obtained in flat Einstein-Straus model. This
limit value is an interesting result that could be regarded as a lower bound
on the radius of the universe today, so that the spatial curvature, which is
a property of the Friedmann universe, doesn't affect the bending of light
nor the time delay.

Furthermore, the results confirm the Rindler and Ishak's claim that a
positive cosmological constant attenuates the bending of light \cite%
{RindlerIsak}.

Even in the case of negatively curved Einstein-Straus model, we expect that
present value of scale factor should also have an effect on the bending of
light and time delay since the spatial curvature only changes its sign ($k=-1
$), leading to a nonzero curvature density as in the case of positively
curved Einstein-Straus model. This case will be investigated in a
forthcoming work. The same problem should also be extended to include the
interior Kottler solution, wherein the photons could pass through the mass
distribution \cite{intK}.

\subsection*{Acknowledgments}

We acknowledge the support of the Ministry of Higher Education and
Scientific Research of Algeria under grant D00920140040.


\begin{thebibliography}{99}
\bibitem{RindlerIsak} Rindler, W., Ishak, M.: The contribution of the
cosmological constant to the relativistic bending of light revisited. Phys.
Rev. D \textbf{76,} 043006, 1-5 (2007). DOI: \href{https://journals.aps.org/prd/abstract/10.1103/PhysRevD.76.043006}%
{10.1103/PhysRevD.76.043006}

\bibitem{sereno} Sereno, M.: On the influence of the cosmological constant
on gravitational lensing in small systems. Phys. Rev. D \textbf{77}, 043004
(2008). DOI: \href{https://journals.aps.org/prd/abstract/10.1103/PhysRevD.77.043004}%
{10.1103/PhysRevD.77.043004}

\bibitem{sereno2} Sereno, M.: The role of Lambda in the cosmological lens
equation. Phys. Rev. Lett. \textbf{102}, 021301, 1-4 (2009). DOI: \href{https://journals.aps.org/prl/abstract/10.1103/PhysRevLett.102.021301}%
{10.1103/PhysRevLett.102.021301}

\bibitem{tschu} Sch\"{u}cker, T.: Cosmological constant and lensing. Gen.
Relativ. Gravit. \textbf{41}, 67--75 (2009). DOI: \href{https://link.springer.com/article/10.1007/s10714-008-0652-2}%
{10.1007/s10714-008-0652-2}

\bibitem{tschu2} Sch\"{u}cker, T.: Strong lensing with positive cosmological
constant. In: Moriond Proceedings Cosmology (2008). \href{https://arxiv.org/abs/0805.1630}%
{arXiv:0805.1630}

\bibitem{mira} Miraghaei, H., Nouri-Zonoz, M.: Classical tests of general
relativity in the Newtonian limit of Schwarzschild-de Sitter spacetime. Gen.
Relativ. Gravit. \textbf{42}, 2947--2956 (2010). DOI: \href{https://link.springer.com/article/10.1007/s10714-010-1052-y}%
{10.1007/s10714-010-1052-y}

\bibitem{ES} Einstein, A., Straus, E. G.: The influence of the expansion of
space on the gravitation fields surrounding the individual star. Rev. Mod.
Phys. \textbf{17}, 120-124 (1946). DOI: \href{https://journals.aps.org/rmp/abstract/10.1103/RevModPhys.17.120}%
{10.1103/RevModPhys.17.120}

\bibitem{Schucking} Sch\"{u}cking, E.: Das Schwarzschildsche Linienelement
und die expansion des Weltalls. Z. Phys. \textbf{137}, 595-603 (1954). DOI: 
\href{https://link.springer.com/article/10.1007/BF01375011}{%
10.1007/BF01375011}

\bibitem{balbinot} Balbinot, R., Bergamini, R., Comastri, A.: Solution of
the Einstein--Straus problem with a $\Lambda $ term. Phys. Rev. D \textbf{38}%
, 2415-2418 (1988). DOI: \href{https://journals.aps.org/prd/abstract/10.1103/PhysRevD.38.2415}%
{10.1103/PhysRevD.38.2415}

\bibitem{ish-rind} Ishak, M., Rindler, W., Dossett, J., Moldenhauer, J.,
Allison, C.: A new independent limit on the cosmological constant/dark
energy from the relativistic bending of light by galaxies and clusters of
galaxies. Mon. Not. R. Astron. Soc. \textbf{388}, 1279--1283 (2008). DOI: 
\href{https://academic.oup.com/mnras/article/388/3/1279/956182}{%
10.1111/j.1365-2966.2008.13468.x}

\bibitem{TSch1} Sch\"{u}cker, T.: Strong lensing in the Einstein-Straus
solution. Gen. Rel. Grav. \textbf{41}, 1595--1610 (2009). DOI: \href{https://link.springer.com/article/10.1007/s10714-008-0731-4}%
{10.1007/s10714-008-0731-4}

\bibitem{kantow} Kantowski, R., Chen, B., Dai, X.: Gravitational lensing
corrections in flat Lambda CDM cosmology. Astrophys.J. \textbf{718,} 913-919
(2010). DOI: \href{http://iopscience.iop.org/article/10.1088/0004-637X/718/2/913}%
{10.1088/0004-637X/718/2/913}

\bibitem{TSCH} Sch\"{u}cker, T.: Lensing in the Einstein-Straus solution. 
\href{https://arxiv.org/abs/1006.3234}{arXiv:1006.3234}

\bibitem{khrip} Khriplovich, I.B., Pomeransky, A.A.: Does cosmological term
influence gravitational lensing? Int. J. Mod. Phys. D \textbf{17}(12),
2255--2259 (2008). DOI: \href{https://www.worldscientific.com/doi/abs/10.1142/S0218271808013832}%
{10.1142/S0218271808013832}

\bibitem{park} Park, M.: Rigorous approach to the gravitational lensing.
Phys. Rev. D \textbf{78}, 023014, 1-6 (2008). DOI: \href{https://journals.aps.org/prd/abstract/10.1103/PhysRevD.78.023014}%
{10.1103/PhysRevD.78.023014}

\bibitem{sim} Simpson, F., Peacock, J.A., Heavens, A. F.: On lensing by a
cosmological constant. Mon. Not. Roy. Astron. Soc. \textbf{402,} 2009
(2010). DOI: \href{https://academic.oup.com/mnras/article/402/3/2009/989698}{%
10.1111/j.1365-2966.2009.16032.x}

\bibitem{Ishak} Ishak, M.: Light deflection, lensing, and time delays from
gravitational potentials and Fermat's principle in the presence of a
cosmological constant. Phys. Rev. D \textbf{78}, 103006, 1-6 (2008). DOI: 
\href{https://journals.aps.org/prd/abstract/10.1103/PhysRevD.78.103006}{%
10.1103/PhysRevD.78.103006}

\bibitem{TD-ES} Boudjemaa, K.-E., Guenouche, M., Zouzou, S.R.: Time delay in
the Einstein-Straus solution. Gen. Rel. Grav. \textbf{43,} 1707-1731 (2011).
DOI: \href{https://link.springer.com/article/10.1007/s10714-011-1152-3}{%
10.1007/s10714-011-1152-3}

\bibitem{chen} Chen, B., Kantowski, R., Dai, X.: Time delay in Swiss cheese
gravitational lensing. Phys. Rev. D \textbf{82}, 043005, 1-5 (2010). DOI: 
\href{https://journals.aps.org/prd/abstract/10.1103/PhysRevD.82.043005}{%
10.1103/PhysRevD.82.043005}

\bibitem{KTD} Sch\"{u}cker, T. Zaimen, N.: Cosmological constant and time
delay. A\&A \textbf{484}, 103-106 (2008). DOI: \href{https://www.aanda.org/articles/aa/abs/2008/22/aa09449-08/aa09449-08.html}%
{10.1051/0004-6361:200809449}

\bibitem{inada} Inada, N., et al.: A gravitationally lensed quasar with
quadruple images separated by 14.62 arcseconds. Nature \textbf{426}, 810-812
(2003). DOI: \href{https://www.nature.com/articles/nature02153}{%
10.1038/nature02153}

\bibitem{oguri} Oguri, M., et al.: Observations and theoretical implications
of the large separation lensed quasar SDSS J1004+4112. Astrophys. J. \textbf{%
605}, 78--97 (2004). DOI: \href{http://iopscience.iop.org/article/10.1086/382221/meta}%
{10.1086/382221}

\bibitem{ota} Ota, N. et al.: Chandra observations of SDSS J1004+4112:
constraints on the lensing cluster and anomalous X-ray Flux Ratios of the
quadruply imaged quasar. Astrophys. J. \textbf{647}, 215-221 (2006). DOI: 
\href{http://iopscience.iop.org/article/10.1086/505385/meta}{10.1086/505385}

\bibitem{fohl} Fohlmeister, J., et al.: The rewards of patience: An 822 day
time delay in the gravitational lens SDSS J1004+4112. Astrophys. J. \textbf{%
676}, 761-766 (2008). DOI: \href{http://iopscience.iop.org/article/10.1086/528789/meta}%
{10.1086/528789}

\bibitem{kawano} Kawano, Y. Oguri, M.: Time delays for the giant quadruple
lensed SDSS J1004+4112: prospects for determining the densityn profile of
the lensing cluster. Publ. Astron. Soc. Jap. \textbf{58}(2), 271--282
(2006). DOI: \href{https://academic.oup.com/pasj/article/58/2/271/1515327}{%
10.1093/pasj/58.2.271}

\bibitem{intK} Sch\"{u}cker, T.: Lensing in an interior Kottler solution.
Gen. Rel. Grav. \textbf{42, }1991-1995 (2010). DOI: \href{https://link.springer.com/article/10.1007/s10714-010-0978-4}%
{10.1007/s10714-010-0978-4}
\end{thebibliography}
\end{document}